\documentclass[5p,sort&compress]{elsarticle}

\usepackage{lineno}

\journal{Cryogenics}

\usepackage{amsmath,amssymb}

\begin{document}

\begin{frontmatter}

\title{Low temperature transport properties of pyrolytic graphite sheet}

\author[crc]{Sachiko Nakamura\corref{mycorrespondingauthor}}
\cortext[mycorrespondingauthor]{Corresponding author}
\ead{snakamura@crc.u-tokyo.ac.jp}
\author[sphys]{Daisuke Miyafuji}
\author[crc]{Takenori Fujii}
\author[sphys]{Tomohiro Matsui}
\author[crc,sphys]{Hiroshi Fukuyama\corref{mycorrespondingauthor}}
\ead{hiroshi@phys.s.u-tokyo.ac.jp}

\address[crc]{Cryogenic Research Center, The University of Tokyo, 2-11-16, Yayoi, Bunkyo-ku, Tokyo 133-0032, Japan}
\address[sphys]{Department of Physics, The University of Tokyo, 7-3-1, Hongo, Bunkyo-ku, Tokyo 133-0033, Japan}

\begin{abstract}
We have made thermal and electrical transport measurements of uncompressed pyrolytic graphite sheet (uPGS), a mass-produced thin graphite sheet with various thicknesses between 10 and 100\,$\mu$m, at temperatures between 2 and 300\,K. 
Compared to exfoliated graphite sheets like Grafoil, uPGS has much higher conductivities by an order of magnitude because of its high crystallinity confirmed by X-ray diffraction and Raman spectroscopy. 
This material is advantageous as a thermal link of light weight in a wide temperature range particularly above 60\,K where the thermal conductivity is much higher than common thermal conductors such as copper and aluminum alloys.
We also found a general relationship between thermal and electrical conductivities in graphite-based materials which have highly anisotropic conductivities. This would be useful to estimate thermal conductance of a cryogenic part made of these materials from its electrical conductance more easily measurable at low temperature. 
\end{abstract}


\end{frontmatter}

\section{Introduction}
Graphite is an allotrope of carbon with layered structure. 
The structure results in strongly anisotropic thermal and electrical transports~\cite{PhysRev.127.694} along with other unique properties such as self-lubrication~\cite{PhysRevLett.92.126101} and high stability up to 4000~K~\cite{JGRB:JGRB3482}. 
With added characteristics of flexibility, graphite sheets are industrial products making use of those properties. 
Grafoil~\cite{Grafoil} is one of the best known commercial products based on this material, and is used in a wide variety of applications, e.g., sealing gaskets, thermal insulators, electrodes, and as a chemical reagent~\cite{Chung1987}. 
It consists of small natural graphite crystals ($=10$--20~nm)~\cite{Takayoshi2009} which are first powdered, then exfoliated at high temperature, and finally rolled under high pressure. 
It is also widely used as an adsorption substrate for basic research of two dimensional physical and chemical properties of adsorbate thin films~\cite{RevModPhys.79.1381} because of its atomically flat surface of microcrystallites and large specific surface area. 
For this purpose, there is another type of exfoliated graphite called ZYX~\cite{ZYX}, synthesized from HOPG (highly oriented pyrolytic graphite) under rather moderate exfoliation and re-compression conditions, with larger platelet size ($=$100--200~nm). 

Recently, a new flexible graphite sheet, pyrolytic graphite sheet (PGS)~\cite{pgs,ADEM:ADEM201300418}, has been invented. 
PGS is a thin graphite sheet of 10--100~$\mu$m in thickness with a single-crystal-like structure, synthesized by heat decomposition of polymeric films. 
Because of its extremely high in-plane thermal conductivity (2--5 times higher than that of copper at room temperature) and low density ($\approx80$\% of aluminum), PGS and its composites are being used for thermal management in electronic devices like smartphones~\cite{ADEM:ADEM201300418}. 
It is potentially useful for cryogenic applications, especially in space engineering.
One recent example is its use in a vibration isolated thermal link for cryocoolers~\cite{McKinley2016174}. 
However, so far only little is known about physical properties of PGS, including thermal transport at cryogenic temperatures.

In this article, we report results of crystal analysis and electrical and thermal transport measurements at 2--300\,K for PGS. 
Here we focused on an uncompressed version of PGS since it has higher crystallinity and thus higher in-plane conductivity than compressed commercial PGS. 
By measuring both electrical and thermal conductivities, we could deduce a general relationship between them for graphite family materials where the standard Wiedemann-Franz law~\cite{ANDP:ANDP18531650802} is not applicable. 
Other characteristics important for application as an adsorption substrate, such as nitrogen adsorption isotherm and real space imaging of morphology with various microscopes, will be published elsewhere~\cite{part2_tbp}.

\section{Pyrolytic Graphite Sheet (PGS)}
Commercial PGS~\cite{pgs} (hereafter, cPGS) is made first by carbonizing a stack of polymer films of a few $\mu$m thick at $T \lesssim 1000$~K, then by graphitizing the resultant foamed carbon precursor at $T \approx3000$~K, and finally by compression (rubbing) which reduces the thickness by 30--50\%. 
Compared to chemical vapor deposition, which is used to synthesize HOPG, this is a convenient mass production method for thin graphite sheets of good crystallinity. 
Previous transmission electron microscopy (TEM) observations~\cite{10012555627} show that the cross-sectional structure of a similar kind of graphite to that used in the present study is laminar of ultra-thin crystalline graphite layers of 6--7~nm thick which corresponds to 16--20 graphenes.
The average lateral size of each layer is determined to be 10--100~$\mu$m from electron channeling contrast  imaging with scanning electron microscope (SEM)~\cite{10031183877}. 
In general, thinner cPGS has higher crystallinity because, in the graphitization process, the liberated gas can escape more easily and the temperature distribution is more uniform. 

The final compression procedure makes cPGS flexible (like paper) so as to be more useful in practical applications. 
However, it may break the lateral crystalline structure on large scales. 
Therefore, in this work, we mainly studied physical properties of uncompressed PGS (hereafter, uPGS) which is produced by exactly the same method as cPGS except the absence of the final compression. 
As a trade off, uPGS is mechanically brittle and inflexible. 
Thus, to shape it precisely, it is recommended to use a punch designed for cutting thin metal films~\cite{nogamigk}. 

The nominal thicknesses of uPGS studied here are 10, 17, 25, and 100~$\mu$m.
We denote them as uPGS-10$\mu$m, for example, in the following.
Their actual thicknesses measured by micrometer are 19$\pm$2, 29.7$\pm$0.6, 56$\pm$3, and 145$\pm$4~$\mu$m, respectively.
For comparison we also studied properties of cPGS-10$\mu$m whose measured thickness is 13$\pm$2~$\mu$m. 

\section{Crystalline structure and defects}
\begin{figure}[htbp]
\centering
\includegraphics[width=0.9\columnwidth]{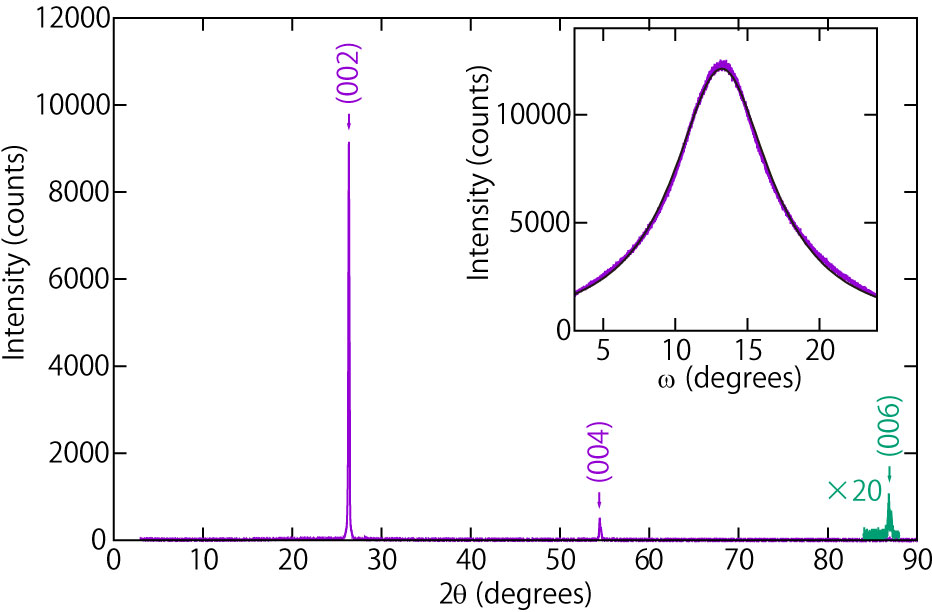}%
\caption{Out-of-plane X-ray diffraction spectrum for uPGS-17$\mu$m. 
(Inset) Rocking curve of (002) peak at $2\theta=26.346$~deg where the FWHM is 8.2~deg.
\label{xrdresults}}%
\end{figure}

Out-of-plane X-ray diffraction was measured for uPGS-17$\mu$m with a powder X-ray diffractometer~\cite{xrd} using Cu K$\alpha_1$ emission. 
The uPGS sample of $10\times10$ mm$^2$ was glued onto a glass holder with GE~7031 varnish. 
Sharp diffraction peaks from graphite are observed indicating that PGS is made purely of graphite crystals (see Fig.~\ref{xrdresults}).
Interplanar spacing $d_{002}$ is determined as $0.33583(7)$~nm from peaks indexed as (002), (004), and (006) using Nelson-Riley function~\cite{doi:10.1080/14786444508520959}.
From the full width at half maximum (FWHM) of the rocking curve of the (002) peak at $2\theta=26.346$~deg, the mosaic angle spread is determined as 8.2$\pm$0.1~deg as shown in the inset of Fig.~\ref{xrdresults}. 
This value is consistent with a mosaic angle spread ($=10\pm$3~deg) roughly estimated from real space SEM imaging of a cross section of uPGS-17$\mu$m~\cite{part2_tbp}.
In-plane X-ray diffraction was also carried out (the data not shown here).
The sample was a stack of 29 uPGS-100$\mu$m sheets (13$\times$5~mm$^2$ each) fixed with epoxy glue (Stycast 1266) each other.
In addition to the peaks from regular spacing between graphene layers such as (002), those from in-plane honeycomb lattice like (110) and peaks indicative of three-dimensional graphite lattice indexed by (101), (102), (103), and (112) are observed.
$d_{002}$ is determined as $0.33592(6)$~nm from the (002), (004), and (006) peaks, and the in-plane lattice parameter $a$ is determined as $0.2463$~nm from the (100) and (110) peaks. 
All these diffraction results agree very well with the previous study for pyrographite films~\cite{doi:10.1063/1.96827}.

Raman spectra of cleaved surfaces of uPGS of 17, 25, and 100~$\mu$m thick were measured at a wavelength of 532~nm using a laser Raman microscope~\cite{raman} with a fixed exposure time of  8~seconds. 
For comparison, cleaved surfaces of HOPG, Grafoil, and ZYX were also measured. 
The most intensive features in Raman spectroscopy for graphite are G ($\approx$1580~cm$^{-1}$) and G' ($\approx$2710~cm$^{-1}$) peaks. 
The relative intensity of G peak to G' one, $I$(G)/$I$(G') is a good representative of the number $n$ of graphene layers for $n \lesssim 6$~\cite{PhysRevLett.97.187401}. 
Measured $I$(G)/$I$(G') values for uPGS are similar to those of other graphites, which confirms that they are thick enough graphites (see Table~\ref{ratio}). 
It is consistent with that all of them have similar FWHM values of G' band ($\approx60$~cm$^{-1}$: not shown in the table). 
D band is known to appear if the surface contains edges or defects where the three-fold symmetry of honeycomb lattice is broken~\cite{rohtua}.
The D band signal was not detected in uPGS and HOPG within experimental errors indicating high crystallinity with immeasurably small amounts of domains and defects~\cite{doi:10.1063/1.369027}. 

\begin{table}[h]
\begin{center}
\caption{Intensity ratios of $I$(G)/$I$(G') and $I$(D)/$I$(G) in Raman spectra for various graphite materials. All the surfaces were cleaved before the measurements.}\label{ratio}
\begin{tabular}{rccc}
\hline \hline
material & nominal thickness & $I$(G)/$I$(G')   & $I$(D)/$I$(G)       \\ \hline
uPGS& $100~\mu$m  & 3.2(1) & $<10^{-3}$ \\
& $25~\mu$m   & 3.1(1) & $<10^{-3}$ \\
& $17~\mu$m  & 3.1(1) & $<10^{-3}$  \\
\\
HOPG    &    ---     & 3.3(2) & $<10^{-3}$   \\
\\
ZYX     &     ---    & 3.2(1) & 0.003(2)   \\
\\
Grafoil & 130~$\mu$m & 3.3(1) & 0.015(4) \\
& 250~$\mu$m & 3.3(1) & 0.033(4)  \\
\hline \hline
\end{tabular}
\end{center}
\end{table}

\section{Electrical resistivity measurement}
We made in-plane ($\rho_{\parallel}$) and out-of-plane ($\rho_{\perp}$) electrical resistivity measurements for uPGS samples of four different thicknesses, i.e., 10, 17, 25, and 100~$\mu$m, in the temperature range between 2 and 300\,K.
They were carried out by the 4-terminal method using the AC transport and resistance options of Physical Properties Measurement System (PPMS) of Quantum Design, Inc.
The typical sample size is $0.5\times9$~mm$^{2}$ for the $\rho_{\parallel}$ measurement and $3\times3$~mm$^{2}$ for the $\rho_{\perp}$ one. 
Gold lead wires of 50~$\mu$m in diameter were glued to the samples with rubber-based carbon paste~\cite{ucc} which adheres strongly to graphite. 

\begin{figure}[htbp]
\centering
\includegraphics[width=0.92\columnwidth]{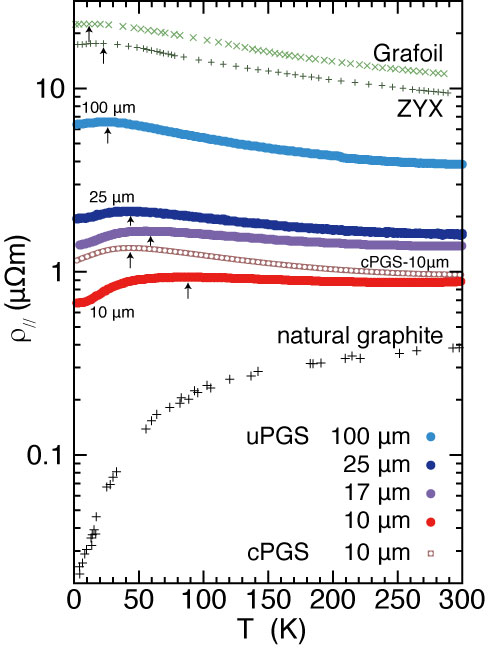}%
\caption{Temperature dependencies of in-plane electrical resistivity $\rho_{\parallel}$ of uPGS with various thicknesses (solid circles) and cPGS-10$\mu$m (open circles). The arrows indicate peak temperatures. Data of Grafoil, ZYX~\cite{niimi:4448} and natural graphite~\cite{PhysRev.95.22} are also plotted.\label{logrhopara}}%
\end{figure}

\begin{figure}[htbp]
\centering
\includegraphics[width=0.97\columnwidth]{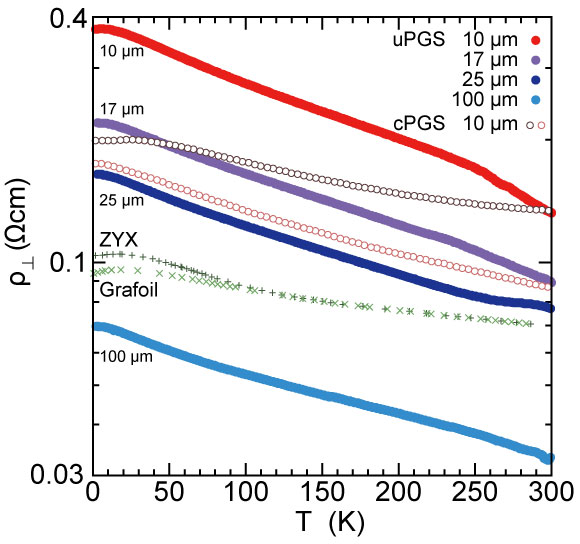}%
\caption{Temperature dependencies of out-of-plane electrical resistivity $\rho_{\perp}$ of uPGS with various thicknesses (solid circles) and cPGS-10$\mu$m (open circles). Data of Grafoil and ZYX are also plotted~\cite{niimi:4448}.\label{logrhoperp}}%
\end{figure}

Results of the $\rho_{\parallel}$ measurement are shown in Fig.~\ref{logrhopara}. 
$\rho_{\parallel}$ values of the uPGS samples are in between those of exfoliated graphites (Grafoil and ZYX) and natural graphite.
Importantly, thinner uPGS has lower $\rho_{\parallel}$ in order of thickness. 
This is consistent with the fact that thinner uPGS has better crystallinity. 
Note that the variation of $\rho_{\parallel}$ with thickness, 3--5 times, is larger than the variation of density, twice at most. 

In exfoliated graphites it is known that there are two different conduction mechanisms.
The first is metallic which dominates the low temperature behavior of $\rho_{\parallel}$ and $\rho_{\perp}$, and the second is variable range hopping (VRH) which dominates high temperature behavior~\cite{PhysRevB.25.4167,niimi:4448}.
As a result, there exists a peak temperature ($T_{\mathrm{peak}}$) in $\rho$ vs. $T$ around 20~K that separates the two behaviors as shown in Fig.~\ref{logrhopara}.
We observed a similar $T$ dependence for uPGS but with higher $T_{\mathrm{peak}}$ (indicated by the arrows in the figure) again in order of thickness. 
As the thickness decreases, the $T$ dependence of $\rho_{\parallel}$ becomes weaker above $T_{\mathrm{peak}}$ and stronger below $T_{\mathrm{peak}}$.
The measured in-plane conductivity $\sigma_{\parallel}$ ($= 1/\rho_{\parallel}(T)$) is well described by the following equation below 150\,K: 
\begin{equation}
\label{sigmaeq}
\sigma=\frac{1}{\rho_{0}+AT}+\sigma_{0}^{\mathrm{h}}\exp\left(-\frac{T_{0}}{T}\right)^{\alpha},
\end{equation}
where the first term corresponds to the metallic channel ($\sigma_{\mathrm{metal}}$) and the second term to the hopping one. 
Within the VRH model,  
\begin{gather*}
T_{0}=\lambda^{3}/D_{\mathrm{F}}k_{\mathrm{B}},\\
\alpha=1/(d+1),
\end{gather*}
for hopping in $d$ spatial dimensions.
Here $\lambda$ and $D_{\mathrm{F}}$ are the decay length of the localized electronic wave function and the density of states at the fermi energy, respectively.
The hopping is presumably between neighboring microcrystallites across domain boundaries.
Thinner uPGS has longer $\lambda$ and smaller $\rho_{0}$, the residual resistivity, presumably because of larger microcrystalline size and less crystalline defects. 
This explains the thickness dependence of $T_{\mathrm{peak}}$.
It is noted that in our analysis $\alpha$ is a fitting parameter unlike the previous works~\cite{PhysRevB.25.4167, niimi:4448}.
The fittings give $\alpha \approx 1.0$, which corresponds to a simple Arrhenius type conduction, for all the samples.

\begin{figure}[htbp]
\includegraphics[width=0.97\columnwidth]{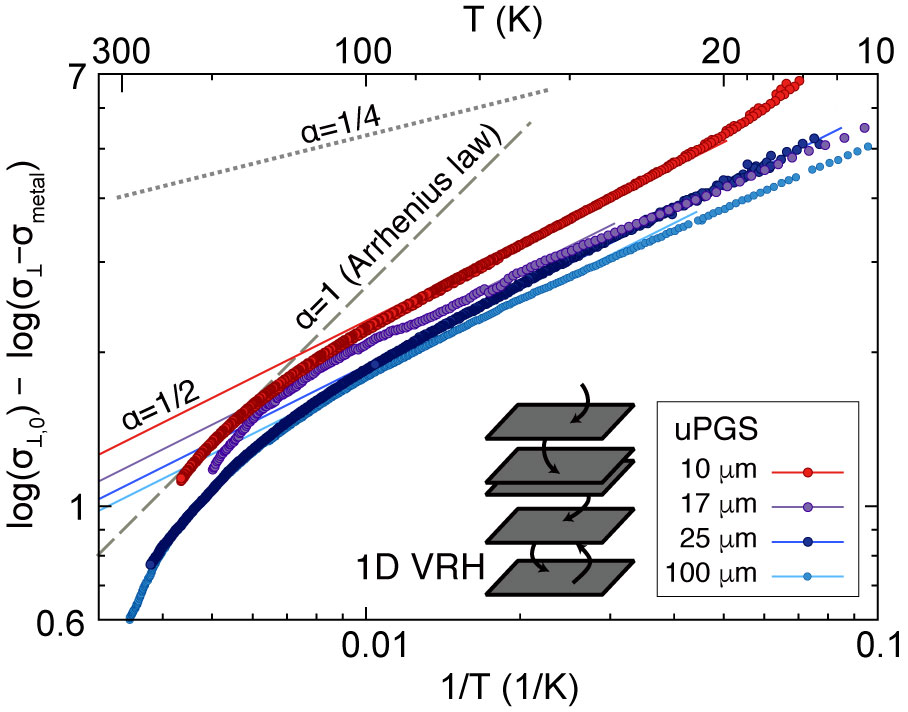}%
\caption{Temperature dependencies of the out-of plane conductivity $\alpha_{\perp}$ of various uPGS samples plotted as logarithm of $\alpha_{\perp}$ vs. $1/T$. 
The data can be fitted to Eq.~\ref{sigmaeq}, a straight line with a slope of $\alpha$ in this plot, much better with $\alpha=1/2$ (thin solid lines) than $\alpha=1/4$ expected from the 3D VRH model (see main text). 
The inset is a schematic diagram of the one dimensional inter-plane hopping.
\label{vrhperp2}}%
\end{figure}

Figure~\ref{logrhoperp} shows results of the $\rho_{\perp}$ measurement.
Again $\rho_{\perp}$ at a fixed temperature varies in order of thickness but with an opposite sign to $\rho_{\parallel}$, i.e., thinner uPGS has larger $\rho_{\perp}$.
This is naturally understood as follows.
Graphite is a layered material with very weak interlayer coupling based on the van der Waals interaction.
Thus the anisotropy of resistivity ($\eta = \rho_{\perp}/\rho_{\parallel} > $$10^2$~\cite{ANDP:ANDP19153531806,PhysRev.95.22}) is so large that it can easily be reduced if the system has a mosaic angle spread or the wavy laminar structure. 
The $\rho_{\perp}$ data  below 150\,K can also be well fitted by Eq.~\ref{sigmaeq} with $\alpha\approx 1/2$, nearly one dimensional hopping, being consistent with the above argument on large $\eta$ (see Fig.~\ref{vrhperp2}).
If we make the same analysis for the previously reported $\rho_{\perp}$ data of Grafoil and ZYX in Refs.~\cite{PhysRevB.25.4167, niimi:4448}, we obtain a similar result ($\alpha=0.42$), although in those papers the data were analyzed assuming $\alpha=1/4$ (three dimensional hopping). 

We note that uPGS of any thickness has a kink in the $T$ dependence of $\rho_{\perp}$ at $T \approx$ 250~K above which the dependence changes randomly in every thermal cycle.
This results in a $\pm$5\% difference in resistivity at 300~K.
The mechanism behind this curious behavior is not known at present. 
The morphology of uPGS with many microscopic inaccessible voids may be changed either by thermal expansion or by desorption/adsorption of gas confined in the voids in every cooling and warming cycle. 
The $T$ dependence of $\rho_{\perp}$ is excellently reproducible at $T < 250$~K for uPGS and in the whole $T$ range we studied for cPGS. 

We comment on effects caused by the final compression to produce cPGS from uPGS from the viewpoint of electrical conductance. 
The open symbols in Figs.~\ref{logrhopara}~and~\ref{logrhoperp} are results of $\rho_{\parallel}$ and $\rho_{\perp}$ measurements for cPGS-10$\mu$m. 
Compared to uPGS-10$\mu$m, the $\rho_{\parallel}$ value at room temperature is only slightly larger.
However, it has a steeper variation down to $T_{\mathrm{peak}}$, and $T_{\mathrm{peak}}$ itself is lower.
Therefore, the overall $T$ dependence of cPGS-10$\mu$m is rather similar to those of the exfoliated graphites.
This would be a result of mixing between $\rho_{\parallel}$ and $\rho_{\perp}$ caused by the compression. 
The same is true for $\rho_{\perp}$ where the cPGS-10$\mu$m samples have much weaker $T$ variations.
This is again similar to the behavior of exfoliated graphite. 
It should be noted that the magnitude of $\rho_{\perp}$ and sometimes even its $T$ dependence differ from sample to sample in the case of cPGS (two different samples are shown in Fig.~\ref{logrhoperp}), presumably because the compression damages the laminar structure.  
Also, $\alpha$ values scatter to a large extent from 0.26 to 0.42. 

\section{Thermal conductivity measurement}
Usually, for metallic samples, it is possible to estimate thermal conductivity ($\kappa_\mathrm{WF}$) from the electrical resistivity $\rho$, which can more easily be measured, through the Wiedemann-Franz (WF) law~\cite{ANDP:ANDP18531650802}: 
\begin{equation}\label{WFeq}
\kappa_{\mathrm{WF}}=\frac{L_0 T}{\rho}, \quad L_0=2.44\times10^{-8}\,\mathrm{W}\Omega\mathrm{K}^{-2}.
\end{equation}
However, in the case of semimetal such as graphite, Eq.~\ref{WFeq} underestimates the true thermal conductivity ($\kappa$) by several orders of magnitude at temperatures where the thermal conduction by phonons plays an important role~\cite{PhysRevB.31.6721}.
Thus we have measured in-plane $\kappa$ of uPGS-10$\mu$m directly using Thermal Transport option of PPMS. 
The sample of 7\,mm wide and 8\,mm long was glued on 4 gold-plated copper electrodes with silver paste.
The electrodes are fixed to a thin rectangular support rod made of Stycast 2850FT ($2\times0.5\times8$\,mm$^{3}$).
The thermal conductance of the support rod is negligibly small at temperatures above 50\,K and is less than half of the total conductance with sample at lower temperatures.
It was carefully measured beforehand and subtracted from the total.

\begin{figure}[h]
\centering
\includegraphics[width=0.97\columnwidth]{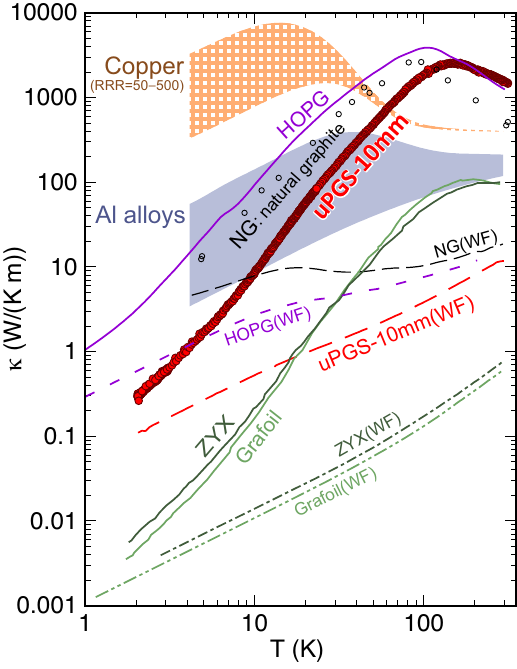}%
\caption{Measured in-plane thermal conductivities $\kappa$ of uPGS-10$\mu$m (this work), HOPG~\cite{PhysRevB.31.6721}, natural graphite~\cite{PhysRev.95.1095}, Grafoil~\cite{UHER1980445}, and ZYX~\cite{UHER1980445}. 
Shadow areas represent $\kappa$ of copper with residual resistance ratios between 50--500 and aluminum alloys of different kinds (Aluminum 1100, 3003-F, 5083-O, 6061-T6, 6063-T5)~\cite{marquardt2002cryogenic}.
Thermal conductivity estimated from electric conductivity using the Wiedemann-Franz law $\kappa_\mathrm{WF}$ is indicated by the broken or dash-dotted lines for each specimen~\cite{PhysRev.95.22,PhysRevB.30.1080,niimi:4448}.  
\label{PGSTT2ai}}%
\end{figure}

In Fig.~\ref{PGSTT2ai}, measured $\kappa$ data of uPGS-10$\mu$m (closed circles) are shown with those of natural graphite (open circles)~\cite{PhysRev.95.1095}, HOPG (solid line)~\cite{PhysRevB.31.6721}, Grafoil~\cite{UHER1980445}, and ZYX ~\cite{UHER1980445}. 
$\kappa$ of uPGS-10$\mu$m is more than one order of magnitude higher than those of ZYX and Grafoil in the whole $T$ range between 2 and 300\,K.
This is a great advantage of this material indicating the longer mean-free path of phonons and thus the higher crystallinity. 
Remarkably, $T$ dependencies of the three kinds of graphite are quite similar to each other.
A peak in $\kappa (T)$ around 150\,K for uPGS-10$\mu$m corresponds to the onset of Umklapp scattering.
It is in between the peak temperature ($T_{\mathrm{peak}} = 100$\,K) of natural graphite and HOPG and those ($T_{\mathrm{peak}} \approx 200$\,K) of Grafoil and ZYX.
At $30 \leq T \leq 100$\,K, the $T$ dependence is $\kappa \propto T^{2.01(2)}$ as expected from two-dimensional phonon conductivity.
It turns to $\kappa \propto T^{2.55(2)}$ at lower temperatures ($2 \leq T \leq 30$\,K) where the conductivity is lower than that of natural graphite by a factor of 4--10. 

Compared to typical metallic thermal conductors, uPGS-10$\mu$m has larger $\kappa$ than copper at $T>60$\,K and than aluminum alloys at $T>40$\,K. 
The $T$ bounds, above which uPGS can transfer heat faster, are extended down to 40\,K and 20\,K, respectively, if we consider the thermal diffusivity $\kappa/C_\mathrm{vol}$ owing to the small density of graphite. 
Here $C_\mathrm{vol}$ is the volumetric specific heat. 
Because the density of graphite is 25\% of copper and 80\% of aluminum, $C_\mathrm{vol}$ is always lower than those metals at any $T$ below 300\,K~\cite{doi:10.1021/ja01623a006,marquardt2002cryogenic}. 
Thus uPGS can be advantageous even at lower $T$ for specific purposes which requires lighter weight and/or smaller heat capacity. 

In Fig.~\ref{PGSTT2ai}, we also plotted $\kappa_{\mathrm{WF}}$ estimated from the measured $\rho_{\parallel}$ through the WF law (broken and dash-dotted lines). 
For all types of the graphite materials, $\kappa$ is much higher than $\kappa_{\mathrm{WF}}$. 
$\kappa/\kappa_{\mathrm{WF}}$ is $\approx$500 at $T\approx100$\,K and slowly decreases with decreasing $T$ down to 3--4 at the lowest temperature. 
In addition, $T$ dependencies of $\kappa_{\mathrm{WF}}$ are quite different from the measured ones.

\begin{figure}[htbp]
\centering
\includegraphics[width=0.97\columnwidth]{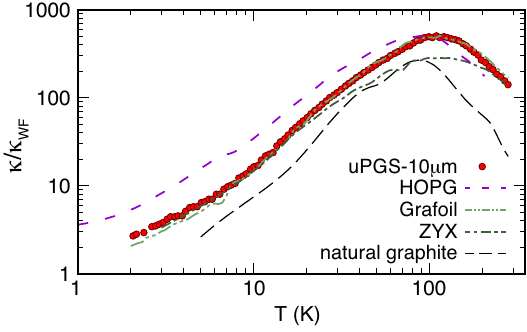}%
\caption{Ratio between the measured in-plane thermal conductivity and that estimated from the measured electrical resistivity through the Wiedemann-Franz law plotted as a function of temperature for uPGS-10$\mu$m, HOPG~\cite{PhysRevB.31.6721,PhysRev.95.22}, natural graphite~\cite{PhysRev.95.1095,PhysRevB.30.1080}, Grafoil~\cite{UHER1980445,niimi:4448}, and ZYX~\cite{UHER1980445}.
\label{ratioplot}}%
\end{figure}

Finally, it is interesting to note that the ratios $\kappa / \kappa_{\mathrm{WF}}$ are rather similar for different graphites, particularly among uPGS-10$\mu$m, Grafoil, and ZYX through the whole $T$ range as shown in Fig.~\ref{ratioplot}.
In this sense, electrical resistance measurement still provides a useful ``rough" estimation of thermal conductance for a variety of graphite materials. 
From the fact, we believe that uPGS-10$\mu$m should have the highest in-plane thermal conductivity among other PGSs though we did not directly measure their $\kappa$. 

\section{Conclusions}
PGS is a recently developed mass-producible pyrolytic graphite sheet with several potential applications including a light-weight and highly conducting thermal link at cryogenic temperatures. 
We made electrical and thermal conductivity measurements at 2$\leq T \leq$300\,K along with X-ray diffraction and Raman spectroscopy characterizations of uncompressed PGSs (uPGSs) of various thicknesses between 10 and 100\,$\mu$m. 
The thinnest uPGS (uPGS-10$\mu$m) has the highest in-plane thermal conductivity ($\kappa$) because of the highest crystallinity. 
By the same reason, uPGS-10$\mu$m and its compressed version (cPGS-10$\mu$m) have more than one order of magnitude higher $\kappa$ than that of Grafoil, a commonly used flexible exfoliated graphite sheet, in the whole temperature range. 
It is a better thermal conductor than copper at $T>$40--60\,K and aluminum alloys at $T>$20--40\,K as natural graphite (NG) crystal and highly oriented pyrolytic graphite (HOPG) are so due to large thermal conduction by phonons. 
Since it is difficult to machine NG and HOPG into arbitrary thickness, uPGS/cPGS have a great advantage over any other materials for application as a thin and inflexible/flexible cryogenic thermal link. 

We also found that there is a general relationship between the thermal conductivity estimated from in-plane electrical conductivity through the Wiedemann-Franz law ($\kappa_\mathrm{WF}$) and $\kappa$ in graphite family materials. 
This is useful so that one can conveniently estimate thermal conductance of a cryogenic part made of those materials from the more easily obtained electrical conductance.

\section{Acknowledgements}
We are grateful to Yoshiya Sakaguchi, Hiroyuki Hase, and Makoto Nagashima of Automotive \& Industrial Systems Company of Panasonic corporation for providing us the uPGS and cPGS samples. 

This work was financially supported by Grant-in-Aid for Scientific Research (B) (Grant No.~15H03684), and Challenging Exploratory Research (Grant No.~15K13398) from JSPS.
The laser Raman microscope used in this work was supplied by MERIT program, The University of Tokyo.

\section*{References}
\bibliography{pgsbib}

\end{document}